\documentclass[twocolumn,prl,showpacs,preprintnumbers,amsmath,amssymb,groupedaddress,floatfix,nofootinbib]{revtex4}
\usepackage{graphicx}
\usepackage{xspace}
\usepackage{hyphenat}
\graphicspath{{../figures/}}

\begin{document}
\newcommand\edem{\emph}
\newcommand\ketz{|0\rangle}
\newcommand\ketm{|-\rangle}
\newcommand\ketp{|+\rangle}
\newcommand\fone{\ensuremath{F\!\!=\!\!1}\xspace}
\newcommand\ftwo{\ensuremath{F\!\!=\!\!2}\xspace}
\title{Spinor Dynamics in an Antiferromagnetic Spin-1 Condensate}
\author{A. T. Black}
\author{E. Gomez}
\author{L. D. Turner}
\author{S. Jung}
\author{P. D. Lett}
\affiliation{Joint Quantum Institute, University of Maryland and\\
National Institute of Standards and Technology, Gaithersburg, Maryland 20899}

\date{\today}

\begin{abstract}
 We observe coherent spin oscillations in an antiferromagnetic spin-1
 Bose-Einstein condensate of sodium. The variation of the spin
 oscillations with magnetic field shows a clear signature of
 nonlinearity, in agreement with theory, which also predicts
 anharmonic oscillations near a critical magnetic field.  Measurements
 of the magnetic phase diagram agree with predictions made in the
 approximation of a single spatial mode. The oscillation period yields
 the best measurement to date of the sodium spin-dependent interaction
 coefficient, determining that the difference between the sodium
 spin-dependent s-wave scattering lengths $a_{f=2}\!-\!a_{f=0}$ is
 $2.47\pm0.27$ Bohr radii.
\end{abstract}


\maketitle

Atomic collisions are essential to the formation of Bose-Einstein
condensates (BEC), redistributing energy during evaporative cooling.
Collisions can be coherent and reversible, leading to diverse
phenomena such as superfluidity~\cite{matthews99} and reversible
formation of molecules~\cite{donley02} in BECs with a single internal
state. When internal degrees of freedom are included (as in spinor
condensates), coherent collisions lead to rich
dynamics~\cite{chang05,kronjager06} in which the population oscillates
between different Zeeman sublevels. We present the first observation
of coherent spin oscillations in a spin-1 condensate with
antiferromagnetic interactions (in which the interaction energy of
colliding spin-aligned atoms is higher than that of spin-antialigned
atoms.)

Spinor condensates have been a fertile area for theoretical studies
of dynamics~\cite{morenocardoner06,law98,ohmi98,zhang05}, ground
state structures~\cite{ho98,zhang03}, and domain
formation~\cite{zhang05b}. Extensive experiments on the
ferromagnetic \fone hyperfine ground state of $^{87}$Rb have
demonstrated spin oscillations and coherent control of spinor
dynamics~\cite{chang05,kronjager05}. Observation of domain formation
in $^{23}$Na demonstrated the antiferromagnetic nature of the \fone
ground state~\cite{stenger98} and detected tunneling across spin
domains~\cite{stamperkurn99}; no spin oscillations have been
reported in sodium BEC until now. The \ftwo state of $^{87}$Rb is
thought to be antiferromagnetic, but a cyclic phase is
possible~\cite{widera06,kuwamoto04}. Experiments on this state have
demonstrated that the amplitude and period of spin oscillations can
be controlled magnetically~\cite{kronjager06}.

At low magnetic fields, spin interactions dominate the dynamics. The
different sign of the spin dependent interaction causes the
antiferromagnetic \fone case to differ from the ferromagnetic one both
in the structure of the ground-state magnetic phase diagram and in the
spinor dynamics. Both cases can exhibit a regime of slow, anharmonic
spin oscillations; however, this behavior is predicted over a wide
range of initial conditions only in the antiferromagnetic
case~\cite{zhang05}. The spin interaction energies in sodium are more
than an order of magnitude larger than in $^{87}$Rb~\fone for a given
condensate density~\cite{chang05}, facilitating studies of spinor
dynamics.

The dynamics of the spin-1 system are much simpler than the spin-2
case~\cite{kronjager06,kuwamoto04,widera06}, having a well-developed
analytic solution~\cite{zhang05}. This solution predicts a divergence
in the oscillation period (not to be confused with the amplitude peak
observed in $^{87}$Rb $\ftwo$~\cite{kronjager06} oscillations).

This Letter reports the first measurement of the ground state magnetic
phase diagram of a spinor condensate, and the first experimental study
of coherent spinor dynamics in an antiferromagnetic spin-1 condensate.
Both show good agreement with the single-spatial-mode
theory~\cite{zhang03}. To study the dynamics, we displace the spinor
from its ground state, observing the resulting oscillations of the
Zeeman populations as a function of applied magnetic field $B$. At low
field the oscillation period is constant, at high field it decreases
rapidly, and at a critical field it displays a resonance-like feature,
all as predicted by theory~\cite{zhang05}.  These measurements have
allowed us to improve by a factor of three the determination of the
sodium \fone spin-dependent interaction strength, which is
proportional to the difference $a_{f=2}-a_{f=0}$ in the spin-dependent
scattering lengths.

The state of the condensate in the single-mode approximation (SMA)
is written as the product $\phi(\mathbf{r})\mathbf{\zeta}$ of a
spin-independent spatial wavefunction $\phi(\mathbf{r})$ and a
spinor
$\zeta=(\sqrt{\rho_{-}}e^{i\theta_{-}},\sqrt{\rho_{0}}e^{i\theta_{0}},
\sqrt{\rho_{+}}e^{i\theta_{+}})$. We use $\rho_{-}$, $\rho_{0}$, and
$\rho_{+}$ ($\theta_{-}$, $\theta_{0}$, and $\theta_{+}$) to denote
fractional populations (phases) of the Zeeman sublevels $m_F=-1$, 0,
and 1, so that $\sum_i \rho_i\!=\!1$. The spinor's ground state and
its nonlinear dynamics may be derived from the spin-dependent part
of the Hamiltonian in the single-mode and mean-field approximations,
subject to the constraints that total atom number $N$ and
magnetization $m\!\equiv\!\rho_{+}\!-\!\rho_{-}$ are
conserved~\cite{zhang05}. The ``classical'' spinor Hamiltonian $E$
is a function of only two canonical variables: the fractional
population $\rho_0$ and the relative phase $\theta \equiv \theta_{+}
+ \theta_{-} -2\theta_{0}$. It is given by
\begin{equation}
    E=\delta(1\!-\!\rho_0)+c\rho_0\!\left(\left(1\!-\!\rho_0\right)
    +\sqrt{\left(1\!-\!\rho_0\right)^2\!-\!m^2}\,\cos{\theta}\right)\!,
    \label{eq:energy}
\end{equation}
where
$\delta\!=\!h\times(2.77\!\times\!10^{10}\textrm{\,Hz}/\textrm{T}^2)B^2$ is
the quadratic Zeeman shift~\cite{zhang05} with $h$ the Planck
constant. (The linear Zeeman shift has no effect on the dynamics.) The
spin-dependent interaction energy is $c\!=\!c_2\left<n\right>$, where
$\left<n\right>$ is the mean particle density of the condensate and
\begin{equation}
    c_2=\frac{4\pi\hbar^2}{3M}(a_{f=2}-a_{f=0})
    \label{eq:c2}
\end{equation}
is the spin-dependent interaction
coefficient~\cite{zhang05,ketterle99}. Here $M$ is the atomic
mass. $a_{f=2}$ and $a_{f=0}$ are the s-wave scattering lengths for a
colliding pair of atoms of total spin $f\!=\!2$ and $f\!=\!0$,
respectively; Bose symmetry ensures there are no s-wave collisions
with total spin of 1.  If $c_2$ is positive (negative), the system is
antiferromagnetic (ferromagnetic). The spinor ground state and spinor
dynamics are determined by Eq.~(\ref{eq:energy}).

The apparatus is similar to that described
previously~\cite{dumke06}. We produce a BEC of $10^5$ $^{23}$Na
atoms in the \fone state, with an unobservably small thermal
fraction, in a crossed-beam 1070\,nm optical dipole trap. The trap
beams lie in the horizontal $xy$ plane, so that the trap curvature
is nearly twice as large along the vertical $z$ axis as in the $xy$
plane. By applying a small magnetic field gradient with the MOT
coils (less than 10\,mT/m) during the 9\,s of forced evaporation, we
fully polarize the BEC: all atoms are in $m_F\!=\!+1$.  Conservation
of spin angular momentum ensures that the magnetization remains
constant once evaporation has ceased; a state with $\rho_{+}\!=\!1$
persists for the lifetime of the condensate, about 14\,s.

We then turn off the gradient field and adiabatically apply a bias
field $B$ of 4 to 51\,$\mu$T along $\hat{x}$, leaving the BEC in the
$\rho_{+}=1$ state. To prepare an initial state, we apply an rf field
resonant with the linear Zeeman splitting; typically the frequency is
tens to hundreds of kilohertz. Rabi flopping in the three-level system
is observed~\cite{sargent76}, and controlling the amplitude and
duration of the pulse can produce any desired magnetization $m$, which
also determines the population $\rho_0$.  The flopping time is less
than 50\,$\mu$s, much shorter than the characteristic times for spin
evolution governed by Eq.~(\ref{eq:energy}). Using this Zeeman
transition avoids populating the $\ftwo$ state, thus avoiding
inelastic losses, which are much greater for $^{23}$Na than for
$^{87}$Rb.

We measure the populations $\rho_i$ of atoms in the three Zeeman sublevels by
Stern-Gerlach separation and absorption imaging~\cite{dumke05}. The
Stern-Gerlach gradient is parallel to the bias field $\vec{B}$, while
the imaging beam propagates in the $\hat{z}$ direction.  The phase
$\theta$ is not measured.

\begin{figure}[h]
  \centering
  \includegraphics[width=3.1in]{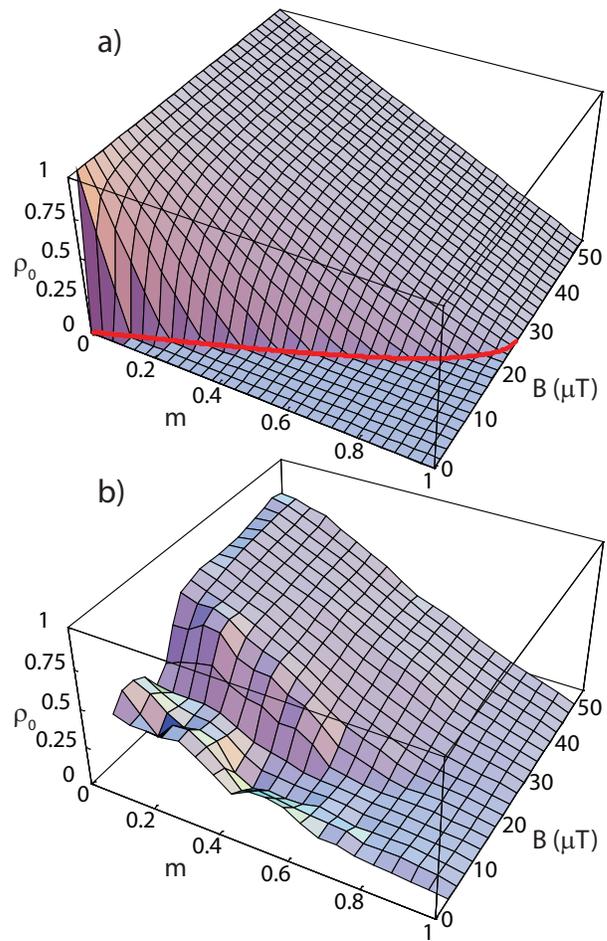}
  \caption{a) Theoretical prediction of the ground-state fractional
  population $\rho_0$ as a function of magnetization $m$ and applied
  magnetic field $B$, assuming a spin-dependent interaction energy
  $c\!=\!h\times$20.5\,Hz. The thick line lying in the $\rho_0=0$
  plane indicates the boundary between the $\rho_0=0$ and the
  $\rho_0>0$ regions. b) Experimental measurement. The surface plot is
  produced by interpolation of data points.  }  \label{fig1}
\end{figure}

To measure the ground state population distribution as a function of
magnetization and magnetic field, we first set the magnetization using the
rf pulse. We then ramp the field to a desired final value
over 1\,s, wait 3\,s for equilibration, and measure the populations
as above.

Figure~\ref{fig1}(b) displays the measured ground-state magnetic phase
diagram. The theoretical prediction in Fig.~\ref{fig1}(a) is the
population $\rho_0$ that minimizes the energy, Eq.(\ref{eq:energy}). Such
minima always occur at $\theta\!=\!\pi$ for antiferromagnetic
interactions. The measurements agree well with the prediction, which
is made for spin interaction energy
$c\!=\!h\!\times\!20.5\,\textrm{Hz}$ (determined by spin dynamics as
described below).

The first term of Eq.~(\ref{eq:energy}) depends on the external
magnetic field and tends to maximize the equilibrium $\rho_{0}$
population. The second, spin dependent, term has the same sign as
$c_2$ and in the antiferromagnetic case tends to minimize the
equilibrium $\rho_{0}$ population. The phase transition indicated by
the thick line in Fig.~\ref{fig1}a arises at the point where these
opposing tendencies cancel for $\rho_0=0$. Along the transition
contour, $\rho_0$ rapidly falls to zero. By contrast, the
ferromagnetic phase diagram has $\rho_0=0$ only at $m=1$. In the
region $B<15$\,$\mu$T and $m>0.6$, there should be virtually no
population in $m_F=0$ for antiferromagnetic interactions, and
populations up to $\rho_0=0.34$ for ferromagnetic interactions
(assuming the same magnitude of $c$).  For our equilibrium data, the
reduced $\chi^2$ with respect to the antiferromagnetic (ferromagnetic)
prediction in this region is 2 (20).  This demonstrates that sodium
\fone spin interactions are antiferromagnetic, as previously shown by
the miscibility of spin domains formed in a quasi-one-dimensional
trap~\cite{stenger98}.

Across most of the phase diagram, the scatter in the population is
consistent with measured shot-to-shot variation in atom number. This
variation is 20\%, implying an 8\% variation in the mean condensate
density according to Thomas-Fermi theory. The variance of results is
not due to the magnetic field (calibrated to a precision of
0.2\,$\mu$T), nor to residual field variations across the BEC (less
than $250$\,pT). Uncertainties in setting the magnetization are
obviated, as the magnetization is measured for each point as the
difference in fractional populations $m\!=\!\rho_{+}\!-\!\rho_{-}$.
Discrepancies between theory and experiment at low magnetic fields may
be attributed to the field dependence of the equilibration time. We
observe equilibration times (see below) ranging from 200\,ms at high
fields to several seconds at low fields, by which time atom loss is
substantial.

If the spinor is driven away from equilibrium, the full coherent
dynamics of the spinor system~Eq.(\ref{eq:energy}) are revealed.  We
initiate the spinor dynamics with the rf transition described above,
but now look at the evolution over millisecond timescales.

The spinor dynamics are described by the Hamilton equations for
Eq.~(\ref{eq:energy})~\cite{zhang05}:
\begin{equation}
\dot{\rho}_0 = -\frac{2}{\hbar} \frac{\partial E}{\partial \theta}
\text{ and } \dot{\theta} =  \frac{2}{\hbar} \frac{\partial
E}{\partial \rho_0}
\label{eq:thetadot}
\end{equation}
The system is closely related to the double-well ``bosonic Josephson
junction'' (BJJ)~\cite{raghavan99,albiez05} and exhibits a regime of
small, harmonic oscillations and, near a critical field $B_c$,  is
predicted to display large, anharmonic oscillations.
At $B_c$ the period diverges (where $\delta(B_c)\!=\!c[(1-\rho_0)+
\sqrt{(1-\rho_0)^2-m^2}\cos{\theta}]$,
with $\rho_0$ and $\theta$ taken at $t=0$)~\cite{zhang05}. The
critical value corresponds to a transition from periodic-phase
solutions of Eq.~(\ref{eq:thetadot}) to running-phase solutions. At
the critical value it is predicted that the population is trapped in a
spin state with $\rho_0\!=\!0$. This phenomenon is related to the
macroscopic quantum self-trapping that has been observed in the
BJJ~\cite{albiez05}. However, very small fluctuations in field or
density will drive $\rho_0$ away from 0.  Observing a ten-fold
increase in the period above its zero-field value would require a
technically challenging magnetic field stability of better than
100\,fT.

\begin{figure}[h]
  \centering \includegraphics[width=3.1in]{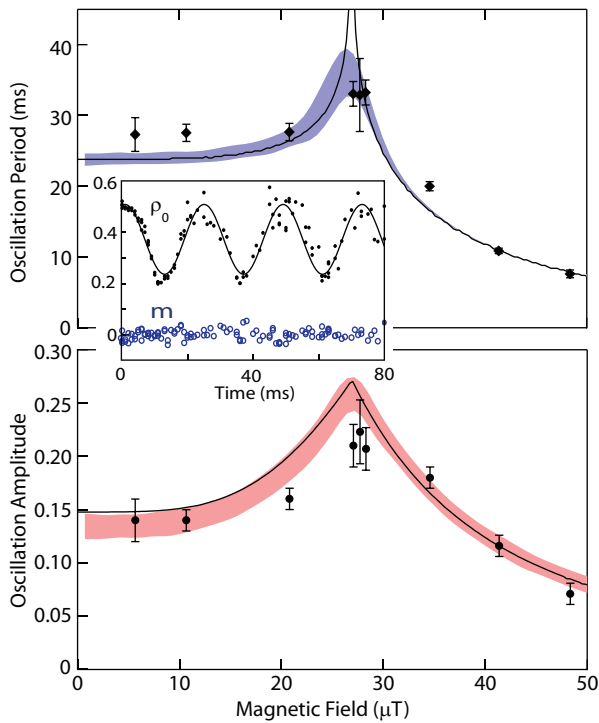}
  \caption{Period (a) and amplitude (b) of spin oscillations as a
  function of applied magnetic field, following a sudden change in
  spin state. The solid lines are theoretical predictions from solving
  Eq.~(\ref{eq:thetadot}). The theoretical prediction of the period
  goes to infinity at about 27\,$\mu$T. The shaded regions are $\pm1$
  standard deviation about the mean values predicted by the Monte
  Carlo simulation.  Inset: Fractional Zeeman population (solid dots)
  and magnetization (open circles) as a function of time after the
  spinor condensate is driven to $\rho_0\!=\!0.5$,
  $m\!=\!0$. $B\!=\!6.1$\,$\mu$T. The solid line is a sinusoidal
  fit.
    } \label{fig:ampper}
\end{figure}

Figure~\ref{fig:ampper} plots the period and amplitude of
oscillation as a function of magnetic field. An example of the
oscillating populations is shown in the inset. The spinor condensate
is prepared with initial $\rho_0\!=\!0.50\pm0.01$~\footnote{All
uncertainties in this paper are one standard deviation combined
statistical and systematic uncertainties} and $m\!=\!0.00\pm0.02$,
and a plot of $\rho_0$ versus time is taken at each field value.
Qualitatively, the period is nearly independent of magnetic field at
low fields, with a small peak at a critical value
$B_c\!=\!28\,\mu$T, followed by a steep decline in period. The
amplitude likewise shows a maximum at $B_c$. Oscillations are
visible over durations of 40\,ms to 300\,ms. Beyond these times, the
amplitude of the shot-to-shot fluctuations in $\rho_0$ is roughly
equal to the harmonic amplitude. This indicates dephasing due to
shot-to-shot variation in oscillation frequency, probably associated
with the variations in magnetic field and condensate density, rather
than any fundamental damping process. At even longer times, we
observe damping and equilibration to a new constant $\rho_0$; the
damping time varies with magnetic field from 200\,ms to 5\,s.

For the theoretical prediction in Fig.~\ref{fig:ampper}, the initial
value of $\rho_0$ and $m$ are obtained experimentally. We treat only
$c$ and $\theta(t\!=\!0)$ as free parameters; $c$ is also predicted by
prior determinations of $c_2$ and our knowledge of the condensate
density. The initial relative phase is \emph{not}
the equilibrium value $\theta\!=\!\pi$, due to our rf preparation. For
a three-level system driven in resonance with both transitions, the
relative phase is $\theta\!=\!0$ at all times during the rf
transition, as we derive from Ref.~\cite{sargent76}. Small deviations
from initial $\theta\!=\!0$ could be caused by an unequal splitting
between the levels, from e.g., the quadratic Zeeman shift.

The best fit to the data in Fig.~\ref{fig:ampper}a~and~b is obtained
by using $c\!=\!h\!\times\!(21\pm2)$\,Hz and
$\theta(t\!=\!0)\!=\!0.5\pm0.3$ (with no other free parameters).
Away from the critical field $B_c$, agreement with theory is good.
The fitted value of $c$ implies that $B_c$ is 27\,$\mu$T, in
reasonable agreement with the apparent peak observed at 28\,$\mu$T.
Our ability to observe strong variations in period near $B_c$ is
limited by density fluctuations (8\%) and magnetic field
fluctuations (0.2\,$\mu$T). Near $B_c$, typically only one cycle is
visible before dephasing is complete. Such rapid dephasing can,
itself, be taken as evidence of a strongly $B$-dependent period, as
expected near the critical field.

To include the known fluctuations in density and magnetic field in our
model, we perform a Monte Carlo simulation of the expected signal,
based on measured, normally distributed shot-to-shot variations in
values of $c$, $\delta$, $m$ and $\rho_0(t\!=\!0)$. At each value of
$B$ in Fig.~\ref{fig:ampper}, we generate 80 simulated time traces,
with each point in the time trace determined from
Eq.~\ref{eq:thetadot}.  We fit the simulated traces using sine waves
and record the mean and standard deviation of the amplitude and period
of the fits. The results (shaded regions in Fig.~\ref{fig:ampper})
show a less sharp peak in the period. The smoothing of the peak at
$B_c$ is consistent with our data.

It is clear in Fig.~\ref{fig:ampper} that the oscillation period is
insensitive to the magnetic field at low values of the field.  In this
regime, the period is sensitive only to the spin interaction $c_2$ and
the density of the condensate $\left<n\right>$.  Measuring this period
allows us to determine the difference in scattering lengths
$a_{f=2}-a_{f=0}$.  The trace inset in Fig.~\ref{fig:ampper} was taken
in this regime, at a magnetic field of $B\!=\!6.1\,\mu$T, and shows
harmonic oscillations with period $24.6\pm0.3$\,ms. Here the predicted
period dependence on magnetic field,
14\,$\mu\mathrm{s}/\mu\mathrm{T}$, is indeed weak and the oscillations
dephase only slightly over the duration shown. Using this measurement
of the period (in which much more data was taken than for each point
making up Fig.~\ref{fig:ampper} (a) and (b)), and including
uncertainties in initial $\theta$, $\rho_0$, and $m$, we obtain the
spin interaction energy $c\!=\!h\times (20.5\pm1.3)\,\textrm{Hz}$.

Finding $a_{f=2}-a_{f=0}$ requires a careful measurement of the
condensate density. We take absorption images with various expansion
times to find the mean field energy. The images yield the column
density in the $xy$ plane, and the distribution in the $z$ direction
can be inferred from our trap beam geometry. We find that the mean
density of the condensate under the conditions of the inset to
Fig.~\ref{fig:ampper} is
$\left<n\right>\!=\!8.6\pm0.9\!\times\!10^{13}$\,cm$^{-3}$.  From
this we calculate $a_{f=2}-a_{f=0}\!=\!(2.47\pm0.27)a_0$, where
$a_0\!=\!52.9$\,pm is the Bohr radius. This is consistent with a
previous measurement, from spin domain structure, of
$a_{f=2}-a_{f=0}\!=\!(3.5\pm1.5) a_0$~\cite{stenger98} and is
smaller than the difference between scattering lengths determined
from molecular levels, $a_{f=2}\!=\!(55.1\pm 1.6) a_0$ and
$a_{f=0}\!=\!(50.0 \pm 1.6) a_0$~\cite{crubellier99}. A multichannel
quantum defect theory calculation gives
$a_{f=2}-a_{f=0}=5.7a_0$~\cite{burke98}.

Finally, we consider the validity of the spatial single-mode
approximation. The SMA was clearly violated in previous work on
$^{23}Na$~\cite{stenger98} and $^{87}Rb$~\cite{chang05} \fone spinor condensates
where spatial domains formed.  Spatial degrees of freedom decouple
from spinor dynamics when the spin healing length $\xi_s\!=\!2\pi
\hbar /\sqrt{2 m |c_2| n}$ is larger than the condensate. From our
density measurements we find typical Thomas-Fermi radii of
(9.4,~6.7,~5.7)\,$\mu$m.
The spin
healing length, based on our measurements of $c$, is typically
$\xi_s\!=\!17$\,$\mu$m. We therefore operate within the range of
validity of the SMA.  Furthermore, Stern-Gerlach absorption images
show three components with identical spatial distributions after
ballistic expansion, indicating that domain formation does not
occur.

In conclusion, we have studied both the ground state and the spinor
dynamics of a sodium \fone spinor condensate. Both agree well with
theoretical predictions in the SMA. By measuring the spin oscillation
frequency at low magnetic field, we have determined the difference in
spin-dependent scattering lengths. The observed peak in oscillation
period as a function of magnetic field demonstrates that the spinor
dynamics are fundamentally nonlinear.  It also suggests the existence
of the predicted regime of highly anharmonic spin oscillations at the
center of this peak, which should be experimentally accessible with
sufficient control of condensate density and magnetic
field. Observation of anharmonic oscillations, as well as population
trapping and spin squeezing effects, could be aided by a minimally
destructive measurement of Zeeman populations~\cite{smith04} to reduce
the effects of magnetic field drifts and shot-to-shot density
variations.

We thank W.~Phillips for helpful discussions, and ONR and NASA for
support. ATB acknowledges an NRC Fellowship. LDT acknowledges an
Australian-American Fulbright Fellowship.

\end{document}